\documentclass[acmlarge]{acmart}
\AtBeginDocument{%
  \providecommand\BibTeX{{%
    \normalfont B\kern-0.5em{\scshape i\kern-0.25em b}\kern-0.8em\TeX}}}

\copyrightyear{2023}
\acmYear{2023}
\setcopyright{acmlicensed}\acmConference[DIS Companion '23]{Designing Interactive Systems Conference}{July 10--14, 2023}{Pittsburgh, PA, USA}
\acmBooktitle{Designing Interactive Systems Conference (DIS Companion '23), July 10--14, 2023, Pittsburgh, PA, USA}
\acmPrice{15.00}
\acmDOI{10.1145/3563703.3596621}
\acmISBN{978-1-4503-9898-5/23/07}

\begin{document}

\title{Decentralized Governance for Virtual Community(DeGov4VC): Optimal Policy Design of Human-plant Symbiosis Co-creation}


\author{Yan Xiang}
\affiliation{%
  \institution{Shanghai Jiao Tong University}
  \city{Shanghai}
  \country{China}}
\email{yanxiang@sjtu.edu.cn}

\author{Qianhui Fan}
\affiliation{%
  \institution{Goldsmiths University of London}
  \city{London}
  \country{United Kingdom}}
\email{qfan001@gold.ac.uk}

\author{Kejiang Qian}
\affiliation{%
  \institution{King's College London}
  \city{London}
  \country{United Kingdom}}
\email{kejiang.qian@kcl.ac.uk}

\author{Jiajie Li}
\affiliation{%
  \institution{Massachusetts Institute of Technology}
  \city{Cambridge}
  \country{United States}}
\email{jiajie@mit.edu}

\author{Yuying Tang}
\affiliation{%
  \institution{Tsinghua University and Politecnico di Milano}
  \city{London and Milan}
  \country{China and Italy}}
\email{tyy21@mails.tsinghua.edu.cn}

\author{Ze Gao}
\authornote{Ze Gao is the corresponding author.}
\affiliation{%
  \institution{Hong Kong University of Science and Technology}
  \city{Hong Kong}
  \country{Hong Kong SAR, China}}
\email{zgapap@connect.ust.hk}

\renewcommand{\shortauthors}{Xiang and Gao, et al.}
\renewcommand{\shorttitle}{Decentralized Governance for Virtual Community(DeGov4VC)}

\begin{abstract}


Does the decentralized nature of user behavior in interactive virtual communities help create rules promoting user engagement? Through scenarios like planting, this framework suggests a new paradigm for mutual influence that allows users to impact communities' political decisions. Sixteen participants in the first round of interviews were involved in the framework's creation. Then we developed and implemented our framework in the community with the help of other stakeholders. This proof-of-concept creates user groups using information from users' daily activities as input and grows the green plants in a virtual environment. Finally, we involved AI agents and stakeholders in the framework test and iterations. Our study's user evaluation of a few key stakeholders demonstrates how our strategy enhances user viscosity and experience. Via human-planting ecosystems in a virtual community, this research gives a fresh viewpoint on decentralized governance and an engaging method for co-creating interactive ecological communities.

\end{abstract}



\begin{teaserfigure}
  \includegraphics[width=\textwidth]{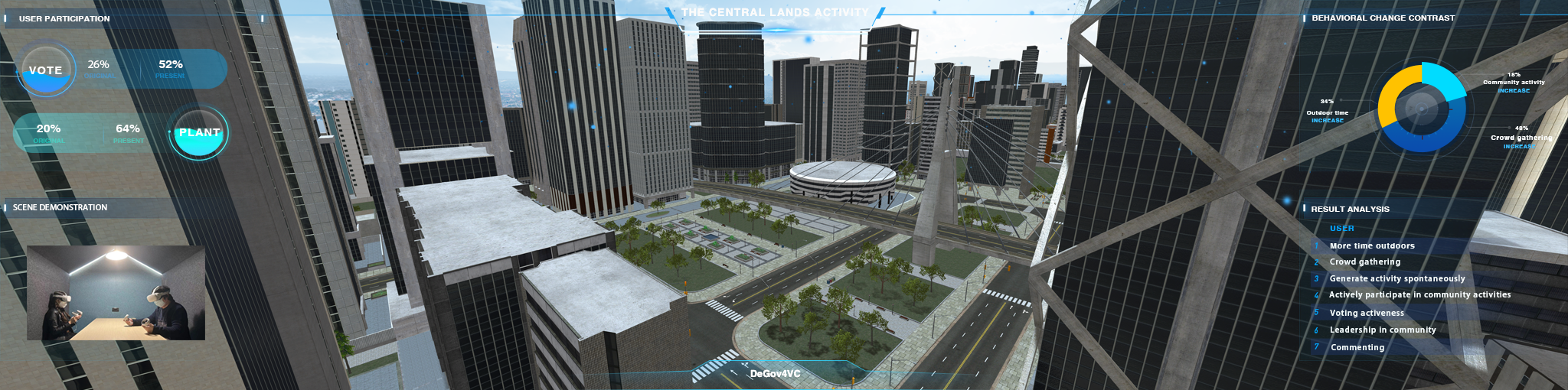}
  \caption{Human-plant Symbiosis Scenario Using the DeGov4VC Platform Interface.}
  \label{fig:teaser}
\end{teaserfigure}

\maketitle

\section{Introduction}
In recent years, how to promote users' interactive co-creation behavior in a VR approach and better engage users in the co-decision-making process in online communities is becoming a focus of attention \cite{za2020value, man2022digital}. New methods of involving users in the interactive co-creation process have proven necessary to ensure the creativity and longevity of communities 
 \cite{clarke2016situated, gao2022bridging}. Many studies have pointed out that this way of co-creation can provide value to participants and communities engaged in the policy-making process \cite{tocchetti2021web, assante2022virtual}. Various scenarios need co-creation with users' participation in developing communities \cite{pillai2020communicate}. Recent HCI research promotes symbiosis with plants as an integral approach to the development of communities \cite{liu2019symbiotic, bettega2021s, fell2020beyond}. If the interactive co-creation processes are to engage and support cooperative symbiosis with nature, we believe they must be placed in such a combination and decision-making process. 

How can users be encouraged to engage in public activities and participate in the decision-making process in the virtual community? Specifically, what gathering forms of decentralized governance in virtual communities can better promote the interactive co-creation process, taking human-plant symbiosis as a scenario, have been significant research problems. 
Thus, our research contributes as follows:
\begin{itemize}
\item Promote a decentralized approach to governance, helping people gain a new concept of decentralized decision-making and a more desirable interactive co-creation community.
\item Create an interactive VR co-creation platform, allowing users to perceive community change through voting. Taking human-plant symbiosis as one of the various scenarios, we emphasize creativity in the design process.
\item Design a framework in the interactive decision-making process through simulation to assist stakeholders with co-creation in the virtual community and evaluate co-design impact through agent-based simulation.
\end{itemize}

\section{Related Works}
Decentralized governance and DAO have received increasing attention in research communities, especially the HCI community. For example, Elsden et al. \cite{elsden2018} arguably identified their vast potential in engaging participants and producing HCI design with blockchain technology. The interactive interface can also facilitate a better understanding of user needs by involving the public in the design decision-making process \cite{yelmi2020}. Besides, Imottesjo and Kain \cite{Imottesjo2022} presented a 3D modeling platform combining web-based VR and mobile augmented reality for urban co-design, emphasizing the necessity of engaging diverse stakeholders and simulating design impact in public space design. An agent-based simulation is an emerging technology to study complex socio-environmental systems and human decision-making, e.g., 3D crowd simulation by Belief–desire–intention(BDI) model \cite{wai2021}.

Although various literature proves that co-design in VR allows remote participants to collaborate in shared virtual environments, some challenging problems, such as incentivizing public engagement, facilitating decision-making, and evaluating co-design impacts, are still pending. Decentralized governance and agent-based simulation can provide a significant chance to address these challenges and actualize the full potential of bottom-up involvement in virtual community design.

\section{Concept and Design Principles}


\subsection{Concept}


The two main focuses of this study are using human-machine collaboration and decentralized governance to facilitate web communities’ participating decision-making and interactive co-creation. This paper artistically and computationally predicts future interfaces co-created by users and AI agents. It then investigates the possibility of using plant growth as a case study in the nature of virtual and physical communities. Users belong to both communities and behavior in the virtual community affects properties in the physical community, while what happens in the physical community is updated in real-time in the virtual community. This highlights the significance of co-creation and co-decision-making to foster interactivity. Our proposed framework, which connects users with AI agents and design framework to construct an independent virtual art community, will be used to create a new virtual interface in the co-creative VR realm.



\subsection{Design Principles}

\begin{figure}[hb]
  \centering
  \includegraphics[height=3cm, width=14cm]{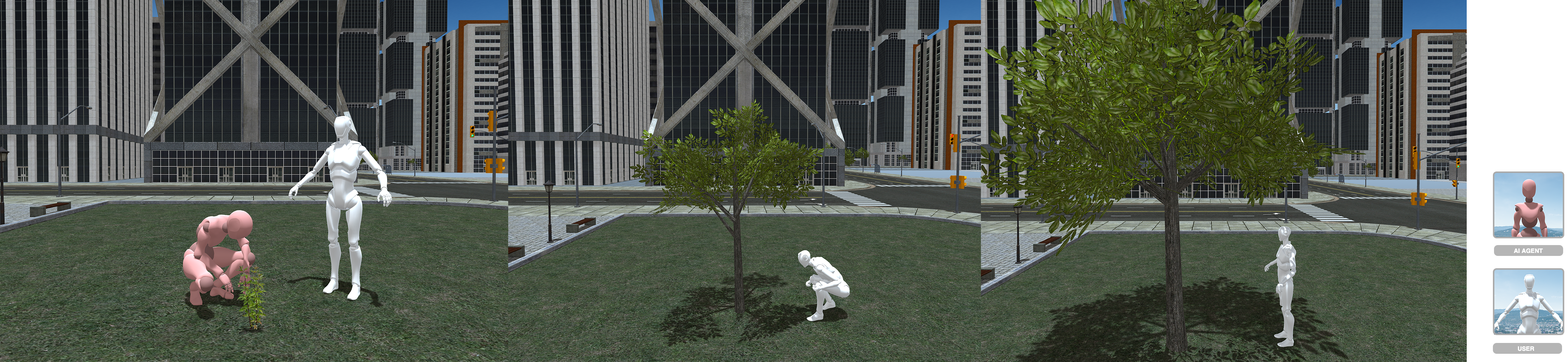}
  \caption{Users or AI Agents Plants in the Virtual Community by Simulating. From Left to Right: Growing Plants From Sapling to a Tree.}
  \label{fig:interaction}
  \vspace{-1.5em} 
\end{figure}

Our design principles for this interactive co-design system are carried out through the following aspects:
\begin{itemize}
\item {\verb|Decentralization|}: Introduce the design philosophy with decentralization in our framework. Unlike most frameworks that use a top-down way, we start from a bottom-up perspective, turning the decision-making process into a collective behavior.
\item{\verb|Scenarios|}: Take the symbiotic scenario of humans and plants as a design case to elicit how to make decisions among the expected behavior between the community and users.
\item{\verb|Co-creation approach|}: Encourage the co-creation of virtual communities by users and AI agents, where users can construct within the virtual community that users and AI agents can co-create in many processes such as planting as shown in Figure \ref{fig:interaction}. Co-creating users and AI agents that help decision-making during the subsequent physical and virtual community creation can result in a more symbiotic community.
\item{\verb|User's experience|}:  Emphasize better human-centered experience of users, through which users' feedback in the decision-making and co-creation process becomes more visible and feasible.
\end{itemize}

\section{Co-design System}

The proposed decentralized governance for the virtual community (DeGov4VC) emphasizes a decentralized approach with virtual reality technology that connects all stakeholders and decisions. This connects many stakeholders in a decentralized way, including community residents, community managers, government, and designers. Such a diverse group of stakeholders had different needs and goals in the system design. Therefore, we adopted an interactive co-design approach \cite{buxton2010sketching} to learn from policy decisions, decentralization, and community residents and their desired experience of human-plant symbiosis. The co-design process focused on two main issues:
\begin{itemize}
\item Identify the shared expectations of multiple stakeholders for a human-plant symbiotic virtual community.
\item Make an interactive decision-making process from multiple opinions and get optimal design results.
\end{itemize}

\begin{table*}[!h]
  \caption{Co-design Workflow}
  \label{tab:commands}
  \begin{tabular}{p{2.5cm}p{3cm}p{2cm}p{6cm}}
    \toprule
    Co-design Phase & Stakeholders & Activity & Aim\\
    \midrule
    \texttt{Phase1:Concept generation} & First round eight participants & User interview & Collect and summary what users expect in our interactive system.\\
    \texttt{Phase2:Framework Creation} & Community,government, and designers & System design & Develop and implement framework in online and offline community.\\
    \texttt{Phase3:Test and Iteration} & Community, AI agents, and government& User evaluation & Verify that our interactive system improves user viscosity and user experience in more scenarios.\\
    \bottomrule
  \end{tabular}
\end{table*}

In addition, to systematically understand the needs of users and managers and design appropriate virtual communities, our research and design path for the DeGov4VC framework is divided into three phases: the concept generation phase, the framework creation phase, and the test and iteration phase. All these three stages can help various stakeholders better participate in our community framework's design and construction process and interactively carry out design iterations and updates in real-time, as shown in Table \ref{tab:commands}, and will be discussed in detail in the discussion section.



\section{Implementation}

\begin{figure}[!h]
  \centering
  \includegraphics[scale=0.4]{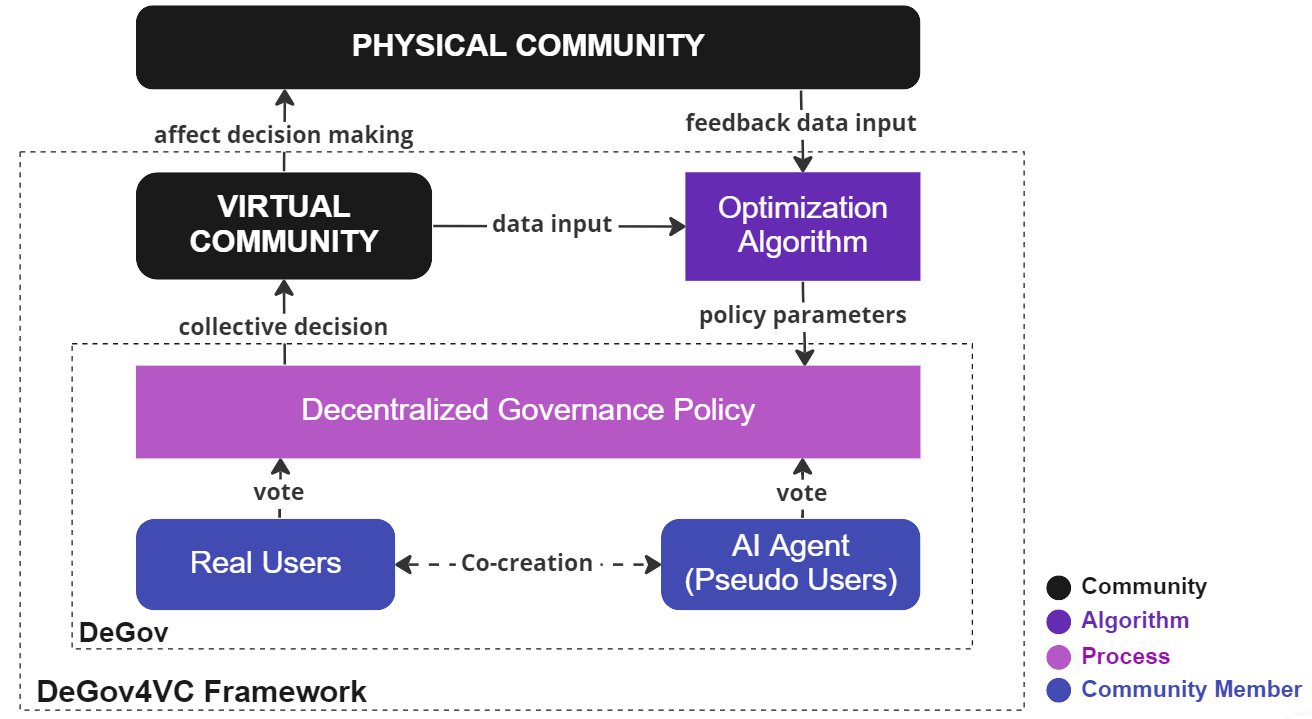}
  \caption{Relationship Diagram of DeGov4VC Framework.}
  \label{fig:relationsh}
  \vspace{-1.5em} 
\end{figure}

In Figure \ref{fig:relationsh}, the DeGov4VC implemented a virtual community with AI agents and provided a client as a user interface. For the client, we will use Unity to develop the front end for map visualization and user interactivity, as shown in Figure \ref{fig:teaser}. Specifically, we will use scan-to-BIM to retrieve the digital twin of a physical community and reproduce its model in Unity. We will also use the GAMA platform \cite{drogoul2013gama} to develop the AI agent. Following the BDI \cite{rao1995bdi, bourgais2020ben} paradigm. We develop and generate AI agent instances to communicate with Unity using the MQTT protocol (e.g., to receive the user's input and send the agent's decision). In the human-machine co-creation process, we use a parameterized decentralized governance policy to regulate the community co-creation process, which determines the community structure and the collective decision-making process. In addition, the outcomes in the virtual community will influence the decision to plant in the physical world.  As decisions are made, we will collect feedback data from virtual (online) and physical (offline) communities to form a dataset and use optimization algorithms to optimize the policy iteratively. 

\begin{figure}[!b]
  \includegraphics[width=\linewidth]{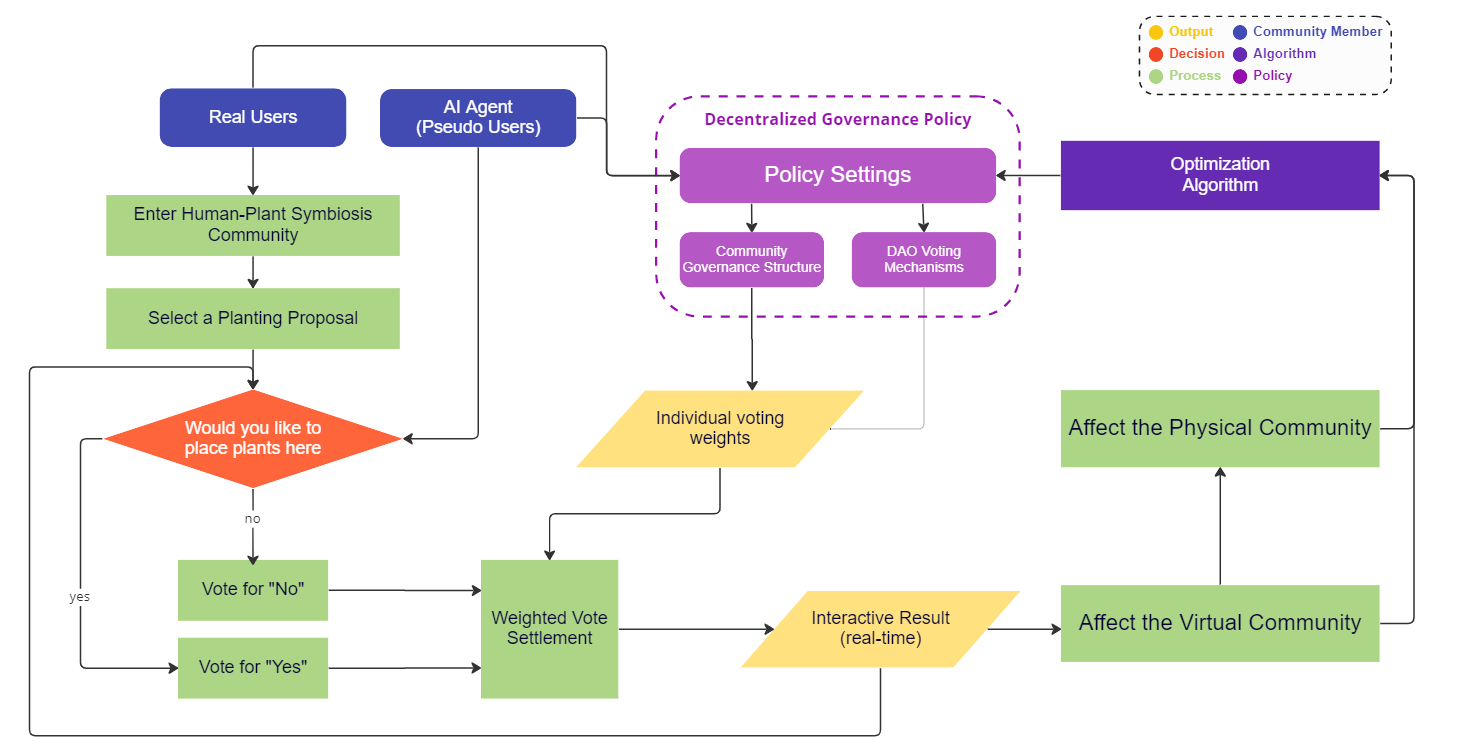}
  \caption{Flowchart of DeGov4VC Framework in Symbiosis Community.}
  \label{fig:flowchart}
  \vspace{-1.5em} 
\end{figure}

\section{Discussion}

This paper introduces a creative and interactive design framework to improve decentralized governance for the virtual community by combining the collaborative creation and policy decisions of real users and AI agents, as can be seen in Figure \ref{fig:flowchart}, which shows the workflow of how different stakeholders will collaborate and interact in our system. The "DeGov4VC" system helps us simulate behavior in virtual reality communities, giving virtual communities more functionality and enabling users to interact more deeply and meaningfully with the VR environment.

Through feedback from co-design participants, we gained valuable insight into how our system caters to different stakeholders. Different communities have well received our framework. This is consistent with past research highlighting the importance of incorporating new technologies into the creative process \cite{seguin2022co, yelmi2020}. The manager also gave positive feedback on how this system can effectively help to make the decision-making process. Finally, residents responded positively to the interactivity of the framework and their ability to engage with the decision-making process in both online and offline communities. However, some audience members suggested that the framework be more open. For instance, encourage and facilitate open communication among members of this interactive community, foster a sense of inclusively by actively seeking out and welcoming diverse perspectives and backgrounds \cite{dow2022scaffolding}, encourage collaboration and sharing of resources among members of the community, and finally, regularly evaluate and update the system and community based on feedback from users, and be transparent about any changes made. Thus, our framework has the potential to be further enhanced to meet the needs and expectations of a broader range of stakeholders.

\section{Conclusion and Future Development}

This research presents a brand-new design framework for user-engaged co-creation in both online and offline communities. DeGov4VC, a co-creation framework, is implemented by our system to create a cooperative symbiosis with nature. The framework investigates a decentralized co-creation framework for community co-creation. Using the example of human-plant symbiosis, this work reveals a more critical perspective on the symbiotic process between broader stakeholders and many community scenarios. Our proposed framework will also assist in developing a new concept of decentralized government and a more desirable network-based co-creative community. As a result, rather than the existing top-down, non-decentralized management style in virtual communities, users will be involved in creating the community (not just in policy decisions related to planting production). Our methodology enhances user experience and participation in virtual communities by including essential stakeholders in the design process.

The effectiveness of a better online community policy in user inventiveness, engagement, and stickiness will be the subject of user studies in future development. We'll conduct surveys, interviews, significance analyses, and semiotic analyses to validate our designs and reveal the participants' feelings about making decisions. Additionally, a semiotic analysis of the visual study of various plant patterns in an online community would be fascinating. Besides, we think that the use of non-fungible tokens in conjunction with user behavior in virtual communities within the metaverse has the potential to serve as a reward and punishment framework. Finally, we would like to investigate further user-user, user-agent, and user-environment interaction techniques in the future. 

\section{Acknowledgement}

This research was partially supported by Tongji University - MIT City Science Lab@Shanghai and the China Academy of Educational Sciences (Project No. JKY16901).

\bibliographystyle{plain}
\bibliography{sample-base}

\end{document}